\documentclass{jfm}

\usepackage{graphicx}
\usepackage{natbib}

\usepackage{amsbsy}
\usepackage{amsfonts}
\usepackage{amsgen}
\usepackage{amssymb}
\usepackage{upmath}
\providecommand\bnabla{\boldsymbol{\nabla}}
\providecommand\bcdot{\boldsymbol{\cdot}}
\title{Eddies and interface deformations induced by optical streaming}
\author[H. Chraibi, R. Wunenburger, D. Lasseux, J. Petit and J-P. Delville]%
{H.\ns C\ls H\ls R\ls A\ls I\ls B\ls I$^1$%
  \thanks{E-mail address: h.chraibi@loma.u-bordeaux1.fr},\ns
R.\ns W\ls U\ls N\ls E\ls N\ls B\ls U\ls R\ls G\ls E\ls R\ls$^1$,\ns D.\ns L\ls A\ls S\ls S\ls E\ls U\ls X\ls$^2$, J.\ns P\ls E\ls T\ls I\ls T\ls$^1$ 
\and J-P.\ns D\ls E\ls L\ls V\ls I\ls L\ls L\ls E\ls$^1$}

\affiliation{$^1$Univ. Bordeaux, LOMA, UMR 5798, F-33400 Talence, France.\\
CNRS, LOMA, UMR 5798, F-33400 Talence, France.\\
$^2$Univ. Bordeaux, I2M, UMR 5295, F-33600 Pessac, France.\\
CNRS, I2M, UMR 5295, F-33600 Pessac, France.\\
Arts et M\'etiers ParisTech, I2M, UMR 5295, F-33600 Pessac, France.\\}

\begin{document}

\maketitle

\begin{abstract}
\large
We study flows and interface deformations produced by the scattering of a laser beam propagating through non absorbing turbid fluids.
Light scattering produces a force density resulting from the transfer of linear momentum
 from the laser to the scatterers. The flow induced in the direction of the beam propagation, called ``optical streaming``, is also able to deform the interface separating the two liquid phases and to produce
 wide humps. The viscous flow taking place in these two liquid layers is solved analytically, in one of the two liquid layers with a stream function formulation,
 as well as numerically in both fluids using a Boundary Integral Element Method.
 Quantitative comparisons are shown between the numerical and analytical flow patterns.
 Moreover, we present predictive simulations dedicated to the effects of the geometry, of the scattering strength and of the viscosities, on both the 
flow pattern and the deformation of the interface. Theoretical arguments are finally put forth to explain the robustness of the emergence of 
secondary flows in a two-layer fluid system. 

\end{abstract}

\section{Introduction}
\begin{figure}
\begin{center}
 \includegraphics[scale=0.4]{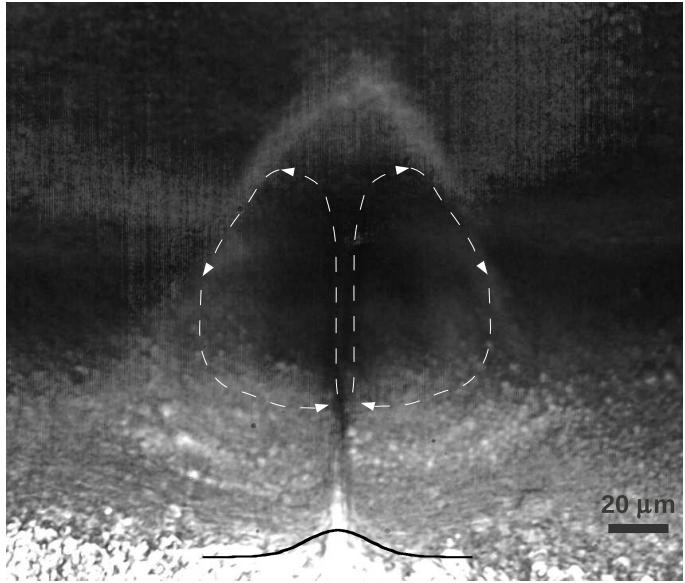}
\caption{Eddies produced by the scattering of a continuous laser beam propagating 
in a non absorbing turbid liquid (micellar phase of microemulsion) layer bounded by a top 
glass wall and a bottom transparent brine solution; a thin layer of metastable 
foam is present at the brine-microemulsion interface. A beam centered toroidal eddy, evidenced by the advection of foam in this picture, induces
 a centered deformation of the interface by viscous 
withdrawal in the direction of beam propagation. The interface deformation is represented by the black curve. The experiment is performed in a so-called Winsor II equilibrium 
\citep{kellay94} obtained here by mixing equal volumes of n-Heptane and brine with a small amount of AOT surfactant \citep{binks00}. 
Sodium chloride is used to screen electrostatic repulsion between surfactant heads and thus
 to significantly reduce the interfacial tension. For the chosen concentrations,
 $[NaCl]=0.05 M$ and $[AOT] = 40 mM$, the interfacial tension is $\sigma=1-5~10^{-6} N/m$. 
The beam power is $P=1.32 W$ and the beam waist $\omega_0 = 2.99\mu m$.
 Eddies are magnified by overlaying 100 frames from a video sequence. }\label{streaming}
\end{center}
\end{figure}

\large 
Streaming flow induced by light scattering has been emphasized by \cite{savchenko97} in various liquids such as liquid crystals and colloids.
 It was showed that a light beam having an
 inhomogeneous transverse intensity profile may enforce convection in liquids as a consequence of the transfer of
 linear and angular momentum from light to liquids.\\
In a recent investigation on the effect of a continuous laser wave on soft liquid-liquid interfaces \citep{schroll07}, we proposed
 an original experiment to optically induce bulk flows in turbid liquids without any heating.
 We showed that light scattering by refractive index fluctuations can produce a steady flow due to transfer
 of momentum from light to the liquid. In addition, the viscous stress 
exerted by this flow can deform the liquid interface.
 We successfully compared experimentally observed deformations of the two-fluid interface to their analytical prediction.
 We call this flow hereafter ''optical streaming`` by analogy with the acoustic streaming phenomenon \citep{nyborg58}
 that occurs when an acoustic wave propagates through a sound absorbing liquid.\\
Indeed, while acoustic streaming is dissipative in nature due to wave absorption, "optical streaming" results from the scattering of the incident
 beam on density inhomogeneities. However, the net result in terms of induced flow is very similar.
More recently, we demonstrated that optical streaming is responsible for the droplet
 emission at the tip of liquid jets triggered by radiation pressure \citep{wunenburger10}.\\
Optical streaming can be observed experimentally as illustrated in Figure \ref{streaming}.\\
\begin{figure}
\begin{center}
 \includegraphics[scale=0.5]{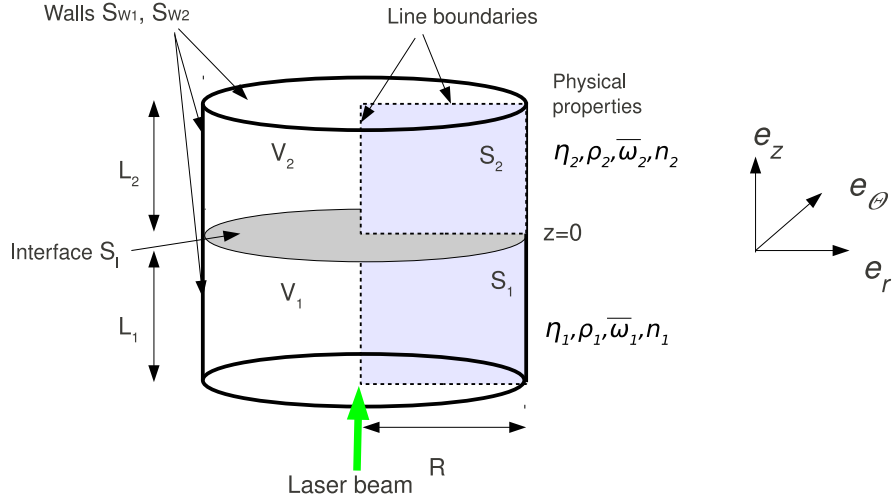}
\caption{Sketch of the flow domain composed of the two fluids and the liquid interface initially at rest.}
\label{numerical}
\end{center}
\end{figure}
In the present investigation, we consider theoretically the optical streaming taking place in a set of two horizontal liquid layers separated by an interface.
 They are submitted to a laser beam propagating upward along the $z$-axis  (see figure \ref{numerical}).
Assuming that at least one of the layers is a turbid liquid, i.e. contains refractive index inhomogeneities smaller that the optical wavelength, a bulk force is induced in this layer by light scattering leading
 to the formation of steady toroidal eddies.
 If the interfacial tension is sufficiently weak, these eddies are able to deform the liquid-liquid interface by the induced viscous stress.
 The goal of the present investigation is thus to study optical streaming in a more general frame, by analyzing the influence of 
the physical properties of the fluids and the geometry of the container. Beyond understanding the nature of these flows, an interesting prospect
 is to use them in future microfluidic experiments \citep{delville09}.\\
Our article is structured as follows: the physical modeling of the
problem is presented in section 2. In section 3, we analytically model the optical streaming
 in a single liquid layer, using a stream function formulation. The numerical simulation in a two layer system  is presented in section 4, using a 
Boundary Element Method. In section 5, comparisons between numerical 
and analytical predictions are discussed, addressing the influence of the aspect ratio and sizes of the cell containing the fluids, of the scattering intensity and of the viscosity ratio on the flow patterns. 
Finally, good quantitative agreements between numerical simulations and experiments are reported in the appendix confirming
 the validity of the physical description and the numerical solution proposed in this work.\\
\section{Formulation of the problem}
\subsection{Geometry and liquid properties}
 We consider two liquid layers labeled by i, i=1,2, separated by an interface, illuminated by a continuous
 laser beam, of Gaussian intensity profile along the radial coordinate and of cylindrical symmetry along the vertical propagation axis $z$. The two liquids are enclosed in a cylindrical container of solid boundaries
 $S_{W1}$ and $S_{W2}$, of radius $R\simeq 10^{-3}m>>\omega_0\simeq 10^{-5}m$ and thicknesses $L_2 \simeq L_1 \simeq 10^{-4}m>>\omega_0$,
 where $\omega_0$ is the characteristic laser beam radius, also called beam waist (see figure \ref{numerical}). 
Considering the axisymmetry of the system along the $z$ axis, we choose cylindrical coordinates (${\boldsymbol e_r}, {\boldsymbol e_\Theta},{\boldsymbol e_z})$
with origin $O$ located at the intersection of the beam axis
with the initially flat interface located therefore at $z=0$.  We refer to a point ${\boldsymbol x}$ by the space coordinates $(r,\Theta,z)$.\\
Physical properties of liquids $i=1,2$ are their viscosities $\eta_i$, densities $\rho_i$, refractive indices $n_i$ and momentum attenuation coefficients $\varpi_i$ which represent the relative 
loss of light linear momentum per unit propagation length. The interfacial tension is denoted
 by $\sigma$. The characteristic values of the liquids are $\eta_i\simeq$ 10$^{-3}$ Pa s, $\rho_i\simeq$10$^3$ kg m$^{-3}$, $n_i\simeq 1.33$, $\varpi_i\simeq 70 m^{-1}$ and $\sigma\simeq 10^{-6} N/m$.
\subsection{Bulk force associated to light scattering}
The momentum flux associated to light of celerity $c$ in vacuum, propagating in a
 fluid of refractive index $n_i$ is $\frac{n_i}{c} I$. The beam intensity profile is 
\begin{equation}
I(r,z)\approx I(r)=\frac{2P}{\upi \omega _{0}^{2}}\mathrm{e}^{-2(r/\omega _{0})^{2}} \label{inten}
\end{equation}
$P$ is the beam power.\\
When light is elastically scattered by the refractive index inhomogeneities of the
 medium, the forward component of its momentum density $q_{iz}$ decreases along
 the direction of propagation. To quantify this decrease, we introduce the forward momentum
 attenuation coefficient in liquid $i$  defined by
\begin{equation}
\varpi_i=- \frac{1}{q_{iz}}\frac{\partial q_{iz}}{\partial z}~~,~~i=1,2
\end{equation}
We assumed statistical isotropy of the distribution of the refractive index inhomogeneities
 within the medium. Therefore, the total rate at which the momentum
 density decreases due to scattering is $\varpi_i \frac{n_i}{c} I{\boldsymbol e_z}~~,~~i=1,2$. As a consequence of momentum conservation, the liquids are subjected to a body force density ${\boldsymbol F_i}$ such as
\begin{equation}
{\boldsymbol F_i}=\varpi_i \frac{n_i}{c} I{\boldsymbol e_z}=F_i\mathrm{e}^{-2(r/\omega _{0})^{2}} {\boldsymbol e_z}~,~~i=1,2 \label{fi}
 \end{equation}
\subsection{Flow equations}
 Since the characteristic order of magnitude  of optically driven flows as measured in the experimental investigation reported in \cite{wunenburger10} is $u_i\simeq$10$^{-4}$ m s$^{-1}$, we investigate a viscous flow at low Reynolds number ($Re_i=u_i L_i \rho_i/\eta_i\simeq 10^ {-2}$).\\ 
Both liquids satisfy the mass conservation and Stokes equations,
\begin{equation}
\bnabla  \bcdot{\boldsymbol u_i} = 0~~;~~  i=1,2
\label{mass}
\end{equation}
\begin{equation}
{\boldsymbol 0} = -\nabla p_i + \eta_i \Delta {\boldsymbol u_i}+{\boldsymbol F_i}~~;~~  i=1,2
\label{stokes}
\end{equation}
The gravitational force has been included in the corrected pressure term $p_i$ that is defined by $p_i=p_i'+\rho_i g z$, $p_i'$ being the fluid pressure.\\
Assuming no slip at the interface and immiscible fluids, the hydrodynamic stress balance on the interface (described by its height $h(r)$)  involving interfacial tension and gravity effects is written as
\begin{equation}
 B.C.1:~~{\boldsymbol T_1}\bcdot{\boldsymbol n}-{\boldsymbol T_2}\bcdot{\boldsymbol n} =(\sigma \kappa -(\rho_1-\rho_2) g h) {\boldsymbol n}~~on~~S_I\\
\label{stressjump}
\end{equation}
Here, $\displaystyle{{\boldsymbol T_i}= -p_{i} {\boldsymbol I} + \eta_i(\bnabla {\boldsymbol u_i} + \bnabla {\boldsymbol u_i}^t)}$  is
 the corrected hydrodynamic stress tensor, $\boldsymbol{n}$ is the unit vector normal to
 the interface directed from fluid 1 to fluid 2 and
$\displaystyle{\kappa(r)=\frac{1}{r}\frac{d}{dr}\frac{r\frac{dh}{dr}}{\sqrt{1+{\frac{dh}{dr}}^2}}}$
 is the double mean curvature of the axisymmetric
 interface in cylindrical coordinates.
The continuity of velocity at the interface yields
\begin{equation}
B.C.2:~~{\boldsymbol u_1} = {\boldsymbol u_2}~~on~~S_I
\end{equation}
 At the container walls, the classical no-slip condition for viscous flow is assumed, i.e. 
\begin{equation}
B.C.3:~~{\boldsymbol u_1}={\boldsymbol 0}~~on~~S_{W1}~~  ;~~B.C.4:~~ {\boldsymbol u_2}={\boldsymbol 0}~~on~~S_{W2}
\end{equation}%
Finally, the motion of the interface is described using a Lagrangian approach 
\begin{equation}
\frac{d{\boldsymbol x}}{dt}={\boldsymbol u}({\boldsymbol x})~~on~~S_I
\label{kin}
\end{equation}%
\section{Analytical solution of a one-fluid model}
When the top fluid is much less viscous and much less dense than the bottom one, a simplified analytical solution 
to the above problem can be derived. Indeed, in this situation of an inviscid top fluid remaining at a constant uniform pressure,
 the stress exerted on the bottom layer can be neglected leading to a one-fluid model.
To develop the solution, the flow taking place in the bottom fluid layer can be solved analytically
 using a stream function formulation.\\
We assume that the width of the laser beam is much smaller than the width of the resulting eddy (in figure \ref{streaming} for instance,
 the eddies presented are ten times larger than the beam waist).
Therefore, the scattering force can be represented as a line force $ f_1{\boldsymbol e_z}=F_1\pi\omega_0^2{\boldsymbol e_z}$ along the centre line of the cylindrical container.\\
The amplitude of the deformation is supposed to be very small compared 
to its characteristic width and a flat interface is assumed.\\

The flow within the layer is modeled as a steady viscous flow contained
in a cylindrical layer of height $L_1$ and radius $R$ centered on the beam axis. 
Due to the symmetry of the optical excitation, the flow is axisymmetric,
 and can be described by a stream function $\psi {\rm (r,z)}$ such that
\begin{equation}
{\boldsymbol u}=\bnabla \times \frac{\psi {\rm (r,z)}}{r}{\boldsymbol e_\Theta } \label{str}
\end{equation}
This choice for the stream function ensures incompressibility (Equation (\ref{mass})), and the Stokes equation becomes
\begin{equation}
 {\boldsymbol 0}=-\nabla p+ \eta_1{\bnabla }^2\left(\bnabla \times \frac{\psi {\rm (r,z)}}{r}{\boldsymbol e_\Theta }\right)+{\boldsymbol F_1}\label{stream}
\end{equation}
Taking the curl of Equation
(\ref{stream}) removes the pressure gradient, and yields
\begin{equation}
 \eta_1 D^4\psi \left({\rm r,z}\right){\rm =}r\left(\bnabla {\rm \times }{\boldsymbol F_1}\right)\bcdot{\boldsymbol e_\Theta }\label{D4f}
\end{equation}
where $D^4\bcdot\equiv D^{2}(D^{2}\bcdot)$ and $D^ 2\equiv\frac{\partial^ 2}{\partial r^2}-\frac{1}{r}\frac{\partial }{\partial r}+\frac{\partial Â²}{\partial z^ 2}$.
Assuming a line force, the right-hand term of Equation (\ref{D4f})  is a delta-function at $r=0$. Thus, except at $r=0$
\begin{equation}
 D^4\psi \left({\rm r,z}\right)=0\label{D4}
\end{equation}
Assuming separation of variables, the general solution is
\begin{equation}
\psi \left({\rm r,z}\right)=G(r){\sin  kz\ }\label{sol}
\end{equation}
where $k = \upi/L_1$.\\
Note that this solution always satisfies $\frac{\partial u_r}{\partial z}(r,z=0) =0$ and $\frac{\partial u_r}{\partial z}(r,z=-L_1)=0$.
 Even though not strictly correct, this
 last condition ensures a straightforward analytical solution.\\
The solution to the problem is finally achieved using four boundary conditions.
We require by symmetry $u_r(0,z)=0$, no flux through the far wall of the container $r = R$, $u_r(R,z) = 0$ and no slip along it  $u_z(R,z)= 0$.\\
Replacing Equation (\ref{sol}) into Equation (\ref{D4}) we find
\begin{eqnarray}
G\left(r\right)=-\frac{U_f}{4\pi k}r\left(K_1\left(kr\right)\int^{kr}_{kR}{I_1\left(x\right)K_1\left(x\right)xdx-I_1\left(kr\right)\int^{kr}_{kR}{{K_1\left(x\right)}^2}}xdx\right)\nonumber\\ 
-\frac{U_f}{4\pi k}rA(k)\left(I_1\left(kr\right)\int^{kr}_{kR}{I_1\left(x\right)K_1\left(x\right)xdx-K_1\left(kr\right)\int^{kr}_{kR}{{I_1\left(x\right)}^2}}xdx\right) \label{G(r)}
\end{eqnarray}
where $A(k)= \frac{\int^{kR}_0{I_1\left(x\right)K_1\left(x\right)xdx}}{\int^{kR}_0{{I_1\left(x\right)}^2}xdx}$,
 $I_1$ and $K_1$ being the first order modified Bessel functions of first and second kind respectively.\\
The last free parameter $U_f$ can now be estimated by matching the flow amplitude and the forcing. This means that the integrated delta-function forcing
must be related to the scattering induced flow. 
More specifically, we take the cross-sectional integral of Equation (\ref{D4f}) at the mid-plane of the liquid
layer, i.e. at $z = -L_1/2$. On the left-hand side, we get
\begin{equation}
 \eta_1 \int D^ 4\psi dS=\eta_1 U_f  \label{D4sin}
\end{equation}
On the right-hand side, combining Equations (\ref{inten}) and (\ref{fi}), we get
\begin{equation}
 \int r\left(\bnabla {\rm \times }{\boldsymbol F_1}\right)\bcdot{\boldsymbol e_\Theta } dS=2\upi\int_0^\infty r^2 \left(-\frac{\partial F_1(r)}{\partial r}\right)dr=2 P \varpi_1 \frac{n_1}{ c}=f_1 \label{D4P}
\end{equation}
Thus, the characteristic velocity of the light-induced bulk flow $U_f$ is
\begin{equation}
 U_f=\frac{f_1}{ \eta_1}=\frac{F_1 \pi \omega_0^2}{\eta_1} \label{y}
\end{equation}
The velocity components are deduced from Equations (\ref{str}) and (\ref{sol})
\begin{equation}
 u_r(r,z)=-\frac{G(r)}{r}k{\cos  kz\ }\label{uz0}
\end{equation}
\begin{eqnarray}
u_z(r,z)=-\frac{U_f}{4\pi}\left[{-K}_0\left(kr\right)\int^{kr}_{kR}{I_1\left(x\right)K_1\left(x\right)xdx-I_0\left(kr\right)\int^{kr}_{kR}{{K_1\left(x\right)}^2}}xdx\right]\sin  kz\nonumber\\ 
-\frac{U_f}{4\pi}A(k)\left(I_0\left(kr\right)\int^{kr}_{kR}{I_1\left(x\right)K_1\left(x\right)xdx+K_0\left(kr\right)\int^{kr}_{kR}{{I_1\left(x\right)}^2}}xdx\right){\sin  kz\ }\nonumber\\
~~ \label{uz}
\end{eqnarray}
Assuming $p(r,z=-L_1/2)=0$, the corrected pressure field can be deduced from equation (\ref{stream}) by noticing that $\frac{\partial p}{\partial r}=-\eta_1\frac{1}{r}\frac{\partial D^ 2\psi}{\partial z}$ and $\frac{\partial p}{\partial z}=\eta_1\frac{1}{r}\frac{\partial D^ 2\psi}{\partial r}$. It yields
\begin{equation}
p(r,z)=\eta_1 \frac{U_f}{4\pi}k\left[K_0\left(kr\right)+A(k){I}_0\left(kr\right)\right]{\cos  kz\ }\label{pre}
\end{equation}
where $I_0$ and $K_0$ are the $0^{th}$ order modified Bessel functions of first and second kind respectively.
The solution of this simplified model allows to express the corrected normal stress at the top of the
fluid layer which is given by the $T_{zz}$ component of the corrected stress tensor
\begin{equation}
{T }_{zz}\left(r,z\right)=-p\left(r,z\right)+2 \eta_1\frac{\partial u_z}{\partial z} \label{def_Tzz}
\end{equation}
Using Equations (\ref{uz}) and (\ref{pre}), we can write
\begin{equation}
{T }_{zz}\left(r,z\right)=-\eta_1 \frac{U_f}{\pi} k S(r){\cos  kz\ }\label{Tzz}
\end{equation}
with
\begin{eqnarray}
S\left(r\right)=\frac{K_0\left(kr\right)}{4}\left(1-2\int^{kr}_{kR}{I_1\left(x\right)K_1\left(x\right)xdx}\right)-\frac{{I}_0\left(kr\right)}{2}\int^{kr}_{kR}{{K_1\left(x\right)}^2}dx\nonumber\\ 
 +\frac{A(k)}{4}\left[I_0\left(kr\right)\left(1+2\int^{kr}_{kR}{I_1\left(x\right)K_1\left(x\right)xdx}\right)+{2K}_0\left(kr\right)\int^{kr}_{kR}{{I_1\left(x\right)}^2}xdx\right]\nonumber\\
~~
\end{eqnarray}
To predict the deformation, the corrected normal stress exerted by the viscous flow
 in the lower layer is balanced with buoyancy and capillarity.
 Since the interface deformation is considered to remain small (i.e. $h'=dh/dr<<1$), the corrected hydrodynamic stress experienced by the interface at
radius $r$ is well approximated by $-T_{zz}(r,0){\boldsymbol e_z}$.
The linearized equilibrium equation of the interface is therefore written in a dimensionless form using $L_1$ as a characteristic length and $\sigma/L_1$ as a characteristic pressure
\begin{equation}
\mathaccent22{h}''(\mathaccent22r)+\frac{\mathaccent22h'(\mathaccent22r)}{\mathaccent22r}- \hbox{Bo}~ \mathaccent22h(\mathaccent22r)+\frac{f_1}{\sigma} S(\mathaccent22r)=0 \label{equi}
\end{equation}
where $Bo=\rho_1 g L_1^2/\sigma$ is the gravitational Bond number and the symbol $\mathaccent22.$ refers to a dimensionless quantity.\\
This equation is solved using a differential algebraic equation solver (DASSL) based on a backward differentiation formula in the interval $[r=\epsilon, r=R]$.
 Note that $\epsilon \neq 0$ since $T_{zz}(0,0)$ diverges due to the line force distribution exerted at $r=0$.
 Since Equation (\ref{equi}) is of second order, we require two boundary conditions for $h(r)$.
We therefore assume $h(r=\epsilon)=h_\epsilon$ and need the condition $dh/dr(r=\epsilon)=h'_\epsilon$, $h_\epsilon$ being our shooting parameter. 
The convergence condition is chosen to be $dh/dr(r=R)=0$ with a tolerance of $0.1\%$. 
Analytical profiles predictions are compared with experimental and numerical results in the appendix.\\
To check the predictibility of the present simplified one-layer analytical model, we also investigate numerically the two-fluid configuration. This is
presented in the following section.
\section{Numerical resolution}
The Boundary Element Method reveals to be very accurate for
solving interfacial flow problems with high resolution as reported
in the analysis of flow involving electric and magnetic fields \citep{sherwood87}, optical radiation pressure \citep{chraibi08b,chraibi10}
 or buoyancy \citep{manga94,koch94}. When used to solve problems involving surface forces like capillarity, radiation
pressure, etc.. and volume forces, which can be written under a conservative form, this 
method allows to reduce the formulation of the problem to boundary integrals only. In the present case, 
the scattering force obviously does not belong to this class, and consequently the integral formulation 
of the problem involves a volume integral term corresponding to the contribution of this force \citep{Occhialini92,Issenmann11}. The integral formulation
is achieved by first making use of the Green's function for the Stokes operator which corresponds to the 
solution of the fundamental Stokes problem \citep{pozrikidis92}.
 The integral equations are presented below under a dimensionless form by choosing $U_\sigma=\sigma/\eta_1$ as a reference
 velocity, $L_1$ as a reference length and $p_{i0}=\eta_i \sigma/(L_1 \eta_1)~i=1,2$ as reference pressures in both fluids.
 Taking into account the boundary conditions on the interface $S_I$
and on the walls $S_{W1}$, $S_{W2}$ (see figure \ref{numerical}), and denoting by ${\boldsymbol n_e}$
the outward unit normal vector to these boundaries, the integral form providing the velocity at any point
 $\mathaccent22{\boldsymbol x}$ in the interior of the volumes $V_1$ and $V_2$, is given by \\
\begin{equation}
\mathaccent22{\boldsymbol {u_i}}(\mathaccent22{\boldsymbol x})=\int_{S_{I}+S_{Wi}}{\boldsymbol U}\bcdot \mathaccent22{\boldsymbol T_i}\bcdot {\boldsymbol n_e}~ dS_y - \int_{S_{I}+S_{Wi}}{\boldsymbol n_e }\bcdot {\boldsymbol K} \bcdot \mathaccent22{\boldsymbol u_i}~  dS_y + \int_{V_{i}} 
{\boldsymbol U} \bcdot \mathaccent22{\boldsymbol F_i}~  dV_y~~i=1,2
\end{equation}
The solution of this equation requires a prior determination of stress on $S_I$, $S_{W1}$ and $S_{W2}$ as well 
as of the velocity on $S_I$. This is achieved by solving an integral equation which, once B.C.1 (equation (\ref{stressjump})) is
explicitly taken into account, can be written as
\begin{eqnarray}
\lefteqn{\frac{1+\zeta}{2}\mathaccent22{\boldsymbol u}(\mathaccent22{\boldsymbol x}) = \int_{S_I} {\boldsymbol
U}\bcdot {\boldsymbol n} \left(\mathaccent22\kappa(\mathaccent22r_y)-\hbox{Bo}~ \mathaccent22h(\mathaccent22r_y)\right)~  dS_y +} \nonumber  \\
& & (\zeta-1)\int_{S_I}{\boldsymbol n} \bcdot {\boldsymbol K} \bcdot\mathaccent22{\boldsymbol u}
~dS_y+\int_{S_{W1}}{\boldsymbol U} \bcdot \mathaccent22{\boldsymbol T_1} \bcdot {\boldsymbol n}~dS_y-\zeta\int_{S_{W2}} {\boldsymbol
U}\bcdot \mathaccent22{\boldsymbol T_2} \bcdot {\boldsymbol n}~dS_y  +\nonumber  \\
& & \int_{V_{1}} {\boldsymbol
{U}}\bcdot \mathaccent22{\boldsymbol F_1}~dV_y +\zeta \int_{V_{2}} {\boldsymbol
{U}}\bcdot \mathaccent22{\boldsymbol F_2}~dV_y \label{zebigone}
\end{eqnarray}
where $\mathaccent22{\boldsymbol x}$ locates a point on $S_I$, ${\boldsymbol n}={\boldsymbol n_{12}}$ and $\zeta = \eta_2/\eta_1$ is the viscosity
 ratio. \\
Here \textbf{U} and \textbf{K} are Green kernels for velocity and
stress respectively and are given by \citep{pozrikidis92}
\begin{eqnarray}
{\boldsymbol U}({\boldsymbol d})&=&\frac{1}{8\upi}(\frac{1}{d}{\boldsymbol I}+\frac{{\boldsymbol d}{\boldsymbol d}}{d^3}),~i=1,2\\
{\boldsymbol K}({\boldsymbol d})&=&-\frac{3}{4\upi}(\frac{{\boldsymbol d}{\boldsymbol d}{\boldsymbol d}}{d^5}),~i=1,2
\end{eqnarray}
where ${\boldsymbol d}=\mathaccent22{\boldsymbol y}-\mathaccent22{\boldsymbol x}$, $\mathaccent22{\boldsymbol y}(r_y,z_y)$ is the integration point.
The first term in the right hand side of equation (\ref{zebigone})
describes the flow contribution from interfacial tension and gravity, whereas the second term accounts
for shear rates contrast on the interface. This term vanishes when
$\zeta=1$. The third and fourth terms account for shear
occurring on $S_{W1}$ and $S_{W2}$ as a result of the no-slip
boundary condition. The two last terms account for the contribution of the flow due to the
scattering force density. Unlike the other terms, they are evaluated on the volumes $V_1$ and $V_2$ respectively for fluid 1 and 2. 
Velocities on the interface as well as stress over all the
boundaries $S_{I}$, $S_{W1}$ and $S_{W2}$ are determined by
solving the discrete form of this equation using a numerical
procedure.
This is performed by an analytical integration in the azimuthal direction \citep{lee82,graziani89} reducing 
all surface boundaries $S_{I}$, $S_{W1}$, $S_{W2}$ into line boundaries $\Gamma_I$, $\Gamma_{W1}$, $\Gamma_{W2}$
and volumes $V_1$ and $V_2$ into surfaces $S_1$ and $S_2$ (see figure \ref{numerical}).
The numerical procedure requires then the discretization of all line boundaries and surfaces with line segments and quadrilaterals respectively.
Line discretization makes use of constant boundary elements, i.e. line segments with centered nodes.
Elliptic line integrals resulting from the azimutal integration are evaluated using Gauss quadratures and 
power series expansions \citep{bakr85}. Numerical integration on quadrilaterals is performed using two-dimensional
 Gauss quadratures \citep{davis84} and isoparametric transformations.
The fluid-fluid interface is parameterized in terms of arc
length and is approximated by local cubic splines, so that the
curvature can be accurately computed. Distribution and number of
points are adapted to the shape of the interface, so that the
concentration of elements is higher in regions where the variation
of curvature of the interface is larger.\\ The motion of the interface is followed using
the kinematic condition (\ref{kin}) which is discretized using an
explicit first-order Euler time scheme. A typical computation
begins with a flat interface at rest. The laser beam is switched
on at t=0, and the deformation of the interface begins. Computation stops when an equilibrium
state is reached ($\displaystyle{\frac{dh}{dt}\rightarrow0}$). 
The time step is chosen to be
about 50 times smaller than the reference time $\tau=L_1/U_\sigma$.
\section{Results}
\subsection{Single layer flow}
In this section, we compare the steady velocity field in the bottom layer predicted using both the numerical
 simulation and the analytical one-layer model. For this comparison, we chose $\eta_2/\eta_1=10^{-3}$ and $\rho_2<<\rho_1$ while $f_2=0$.
Moreover, the influences of the layer aspect ratio  $R/L_1$ and of the layer thickness to beam size ratio $L_1/\omega_0$ on the flow field and interface shape are discussed.
\subsubsection{Flow pattern}
\begin{figure}
\begin{center}
 \includegraphics[scale=0.68]{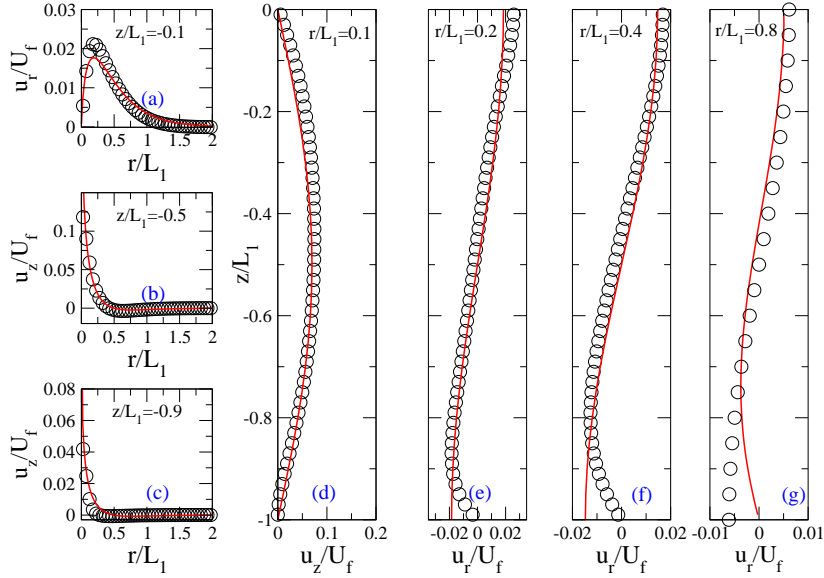}
\caption{Comparison between numerical (symbols) and analytical (lines) velocity fields in the bottom layer.
The left figures show the variation of the reduced  velocity components $u_r/U_f$ and $u_z/U_f$ with respect to $r/L_1$ ($U_f=f_1/\eta_1$).
 The right ones present the variation of  $u_r/U_f$ and $u_z/U_f$ with respect to $z/L_1$.
 $f_1/\sigma=0.157,f_2=0$, $\eta_2/\eta_1=10^{-3}$, $R/L_1=5$, $L_1=L_2$ and $L_1/\omega_0=10$.
 In the numerical resolution $Bo=5$ and therefore $h(r=0)/L_1=0.008$.}\label{velocity}
\end{center}
\end{figure}

Comparisons between the flow patterns predicted analytically and numerically are reported
 in figure \ref{velocity}. 
It shows the $r$ and $z$ dependences of the axial (figures \ref{velocity}(\textit{b--d}))
 and radial velocity components (figures \ref{velocity}(\textit{a,e--g})) in the bottom layer.\\
A satisfactory agreement is observed concerning $u_z$ 
(figures \ref{velocity}(\textit{b--d})) whatever the value of $z$.\\
The same remark holds for $u_r$ (figures \ref{velocity}(\textit{a,e--g})) at the horizontal mid-plane of the layer ($z=-L_1/2$) 
and at the fluid-fluid interface ($z=0$). A discrepancy can be identified at the bottom wall ($z=-L_1$) due to the no-slip condition ($u_r(z=-L_1)=0$) 
in the simulation while a free slip condition was chosen in the one-fluid model for simplicity purpose. For all radial velocities we 
can see that $u_r>0$ for $z>-L_1/2$ and $u_r<0$ for $z<-L_1/2$, which is consistent with the statement of a toroidal eddy centered vertically on the mid-plane 
of the fluid layer.\\
Even though the one-layer model is a simplified version of the real situation, the comparison shows that this analytical model captures the essential trends
 of the flow. In the next section, we investigate the effects of the aspect ratio $R/L_1$.
\newpage
\subsubsection{Influence of the container aspect ratio}
\begin{figure}
\begin{center}
 \includegraphics[scale=0.75]{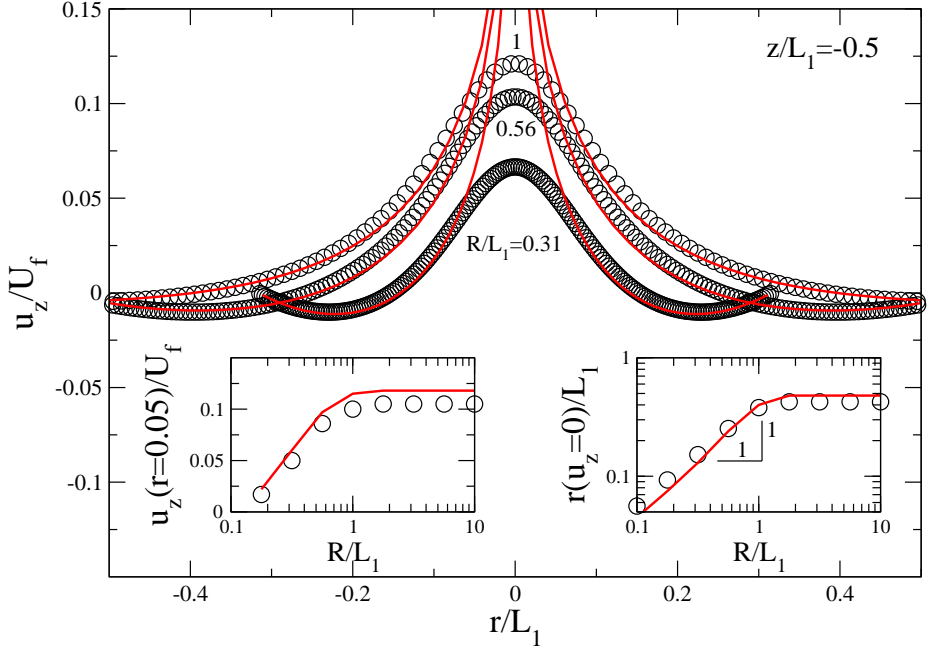}
\caption{Axial velocity for several values of $R/L_1$.
 $f_1/\sigma=0.157,f_2=0$, $\eta_2/\eta_1=10^{-3}$, $L_1=L_2$ and $L_1/\omega_0=10$.
 In the numerical simulation $Bo=5$. Open symbols represent
 the numerical simulation while continuous lines represent the analytical resolution.
 Left inset shows the variation of $u_z(r=0.05)/U_f$ versus $R/L_1$.
 Right inset shows the variation of $r(u_z=0)/L_1$ versus $R/L_1$. }\label{vit-R-L}
\end{center}
\end{figure}

\begin{figure}
\begin{center}
 \includegraphics[scale=0.75]{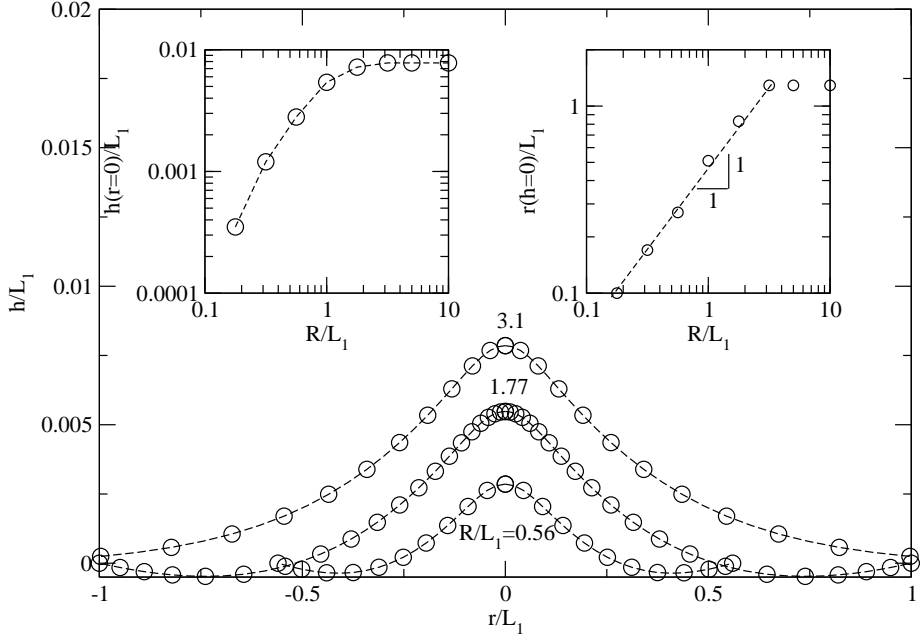}
\caption{Numerical deformation profiles for several values of $R/L_1$.
 $f_1/\sigma=0.157,f_2=0$, $\eta_2/\eta_1=10^{-3}$, $Bo=5$, $L_1=L_2$ and $L_1/\omega_0=10$.
 Left inset shows the variation of $h(r=0)/L_1$ versus $R/L_1$.
 Right inset shows the variation of $r(h=0)/L_1$ versus $R/L_1$.}\label{int-R-L}
\end{center}
\end{figure}

Comparisons between the steady axial velocity profiles $u_z$, predicted analytically and numerically, for different values of the aspect ratio $R/L_1$, are reported in figure \ref{vit-R-L}.
It shows a good agreement between the simulations and the analytical model except for $ |r|/L_1\le0.1$ i.e. within the beam $|r|\le \omega_0$, since $L_1/\omega_0=10$.
This result was expected since the analytical solution shows a divergence of $u_z$ for $r=0$ as a consequence of the line force distribution approximation.\\
The left inset of figure \ref{vit-R-L} shows that $u_z$ scales like $U_f \log(R/L_1)$ for narrow tubes ($L_1 >> R$) while it scales like $U_f$ in the thin films case ($R>>L_1$).
The right inset of figure \ref{vit-R-L} shows that the radial extension of the flow $r(u_z=0)$ scales like the smallest dimension of the container.
In fact,  when $L_1 >> R$, the flow is bounded by the lateral walls and consequently its radial extension scales like $R$.
 In the other limit, when $R>>L_1$, the flow is unbounded in the $r$ direction and mass conservation in 
the axisymmetric geometry produces toroidal eddies of circular section of radius $L_1$.
Deformation profiles predicted by the numerical simulations for different values of $R/L_1$ are presented in figure \ref{int-R-L}.
The left inset shows that the tip deformation scales like $L_1$ when $R>>L_1$.
 Similarly to the velocity field, the radial extension of the deformation scales like the smallest container dimension as shown by the right inset of figure \ref{int-R-L}.
To conclude, in the limit of narrow beams ($L_1>>\omega_0$), the radial extensions of the flow and the deformation always scales like the smallest dimension of the layer.
In the next section, the influence of the beam size on the flow and on the interface deformation is investigated.
\subsubsection{Influence of the beam size}
\begin{figure}
\begin{center}
 \includegraphics[scale=0.75]{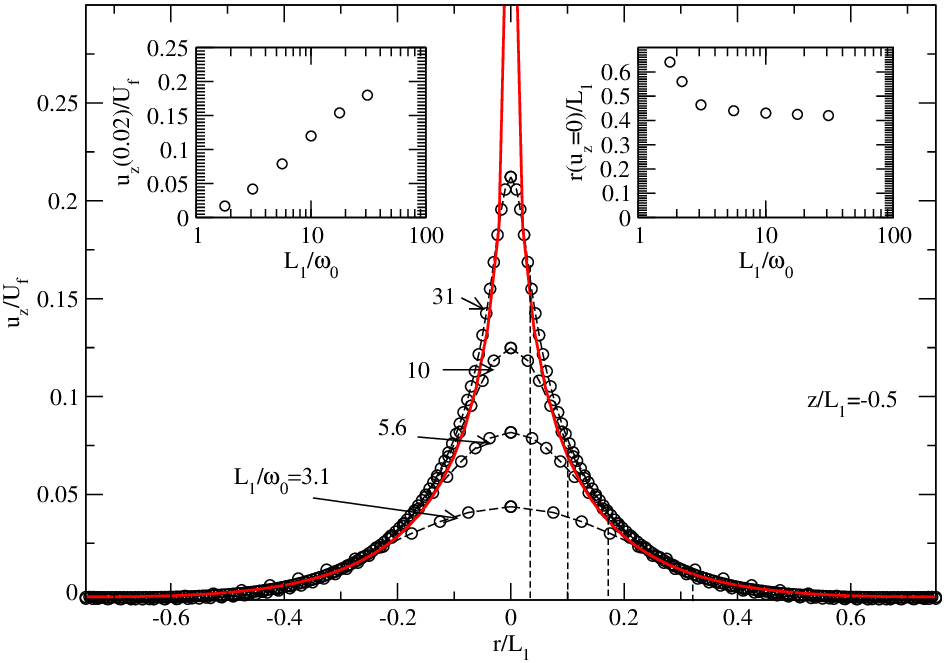}
\caption{Axial velocity for several values of $L_1/\omega_0$.
 $f_1/\sigma=1.57,f_2=0$, $\eta_2/\eta_1=10^{-3}$, $L_1=L_2$ and $R/L_1=5$.
 In the numerical simulation $Bo=5$. Open symbols represent
 the numerical simulation while the continuous line represents the analytical resolution.
 Left inset shows the variation of $u_z(r=0.02)/U_f$ versus $L_1/\omega_0$.
 Right inset shows the variation of $r(u_z=0)/L_1$ versus $L_1/\omega_0$. Beam waist location is shown by the dashed lines for each profile.}\label{vit-L-w0}
\end{center}
\end{figure}

\begin{figure}
\begin{center}
 \includegraphics[scale=0.65]{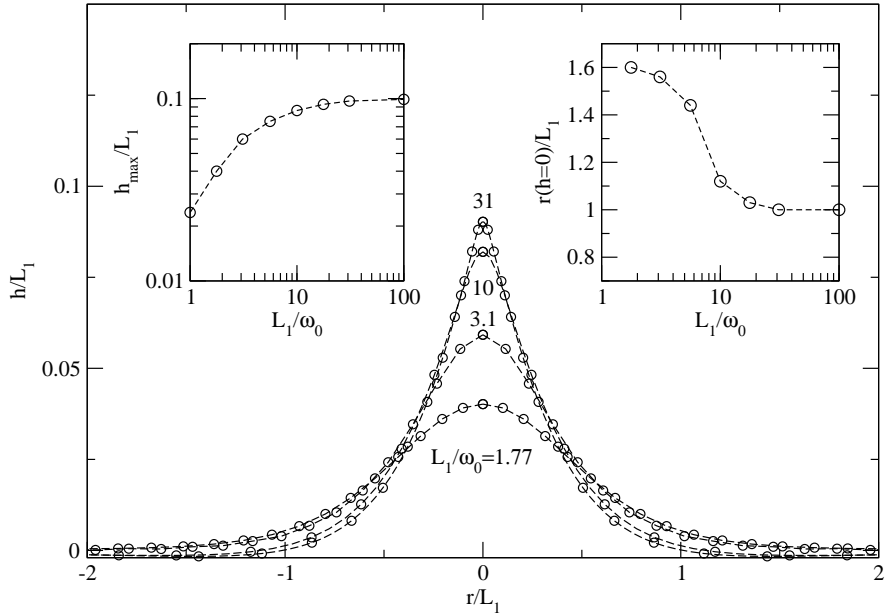}
\caption{Numerical deformation profiles for several values of $L_1/\omega_0$.
 $f_1/\sigma=1.57,f_2=0$, $\eta_2/\eta_1=10^{-3}$, $Bo=5$, $L_1=L_2$ and $R/L_1=5$.
 Left inset shows the variation of the tip height $h(r=0)/L_1$ versus $L_1/\omega_0$.
 Right inset shows the variation of the radial extension of the deformation $r(h=0)/L_1$ versus $L_1/\omega_0$.}\label{int-L-w0}
\end{center}
\end{figure}

Comparisons between the steady axial velocity profiles $u_z$, predicted analytically and numerically, for different values of the layer thickness to beam waist ratio $L_1/\omega_0$, are reported in figure \ref{vit-L-w0}.
It shall be noted that the analytical profile is unique since the solution does not depend on $\omega_0$. We observe that all numerical profiles coincide for $|r|>\omega_0$. In addition, we notice a disagreement for $|r|<\omega_0$, similarly to the results presented in figure \ref{vit-R-L}.
 While the analytical prediction shows a divergence of $u_z$ as $r\rightarrow 0$, the numerical simulation shows a saturation. This saturation always occurs at $r\simeq \omega_0$. We conclude, in the limit of thin films, that there are two characteristic dimensions for the axial velocity.
When $|r|<\omega_0$, the axial velocity depends on $\omega_0$ while its radial extension depends on the layer thickness $L_1$ (see right inset of figure \ref{vit-L-w0}). 
The left inset of figure \ref{vit-L-w0} shows that $u_z$ scales like $U_f \log(L_1/\omega_0)$ in the limit of thin films and narrow beams.  A similar behaviour has been observed, in the $Re<<1$ limit,
 in a theoretical investigation of the acoustic streaming velocity field by Nyborg \citep{hamilton98}.
Indeed, it was demonstrated that the maximum axial velocity of acoustic streaming increases logarithmically with the focal length (equivalent to $L_1$) of the acoustic beam propagating through an infinite fluid.\\
Deformation profiles predicted by the numerical simulations for different values of $L_1/\omega_0$ are shown in figure \ref{int-L-w0}.
We conclude from the insets that both the deformation height and its radial extension scale like $L_1$ when $L_1>>\omega_0$.
The general conclusion of this section is that for thin films $R>>L_1$ and of narrow beams $L_1>>\omega_0$, both the radial extension of the axial velocity
 and of the deformation scale like the layer thickness $L_1$ while their amplitudes scale respectively like
$U_f$ and $L_1$. These limits correspond to our experimental investigations reported previously \citep{schroll07,wunenburger10}.\\
The following sections present the hydrodynamic interaction between the two layers with an investigation of the influence of the different aspect ratios $R/L_1$, $L_1/L_2$ and force and viscosity relative amplitude $f_1/f_2$ and $\eta_2/\eta_1$.
\newpage
\subsection{Double-layer flow}
\subsubsection{Flow patterns}
\begin{figure}
\begin{center}
 \includegraphics[scale=0.3]{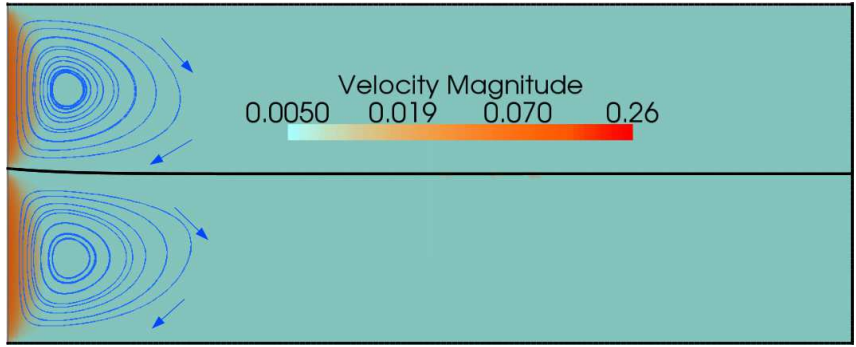}
\caption{Steady flow pattern for $R/L_1=5$ and $L_1/\omega_0=L_2/\omega_0=30$, $\Sigma=f_1/f_2=1$, $\zeta=\eta_2/\eta_1=1$, $Bo=10$ and $f_1/\sigma=0.3$. Velocity magnitude is reduced by $U_\sigma=\sigma/\eta_1$.}\label{2fluids}
\end{center}
\end{figure}

\begin{figure}
\begin{center}
 \includegraphics[scale=0.42]{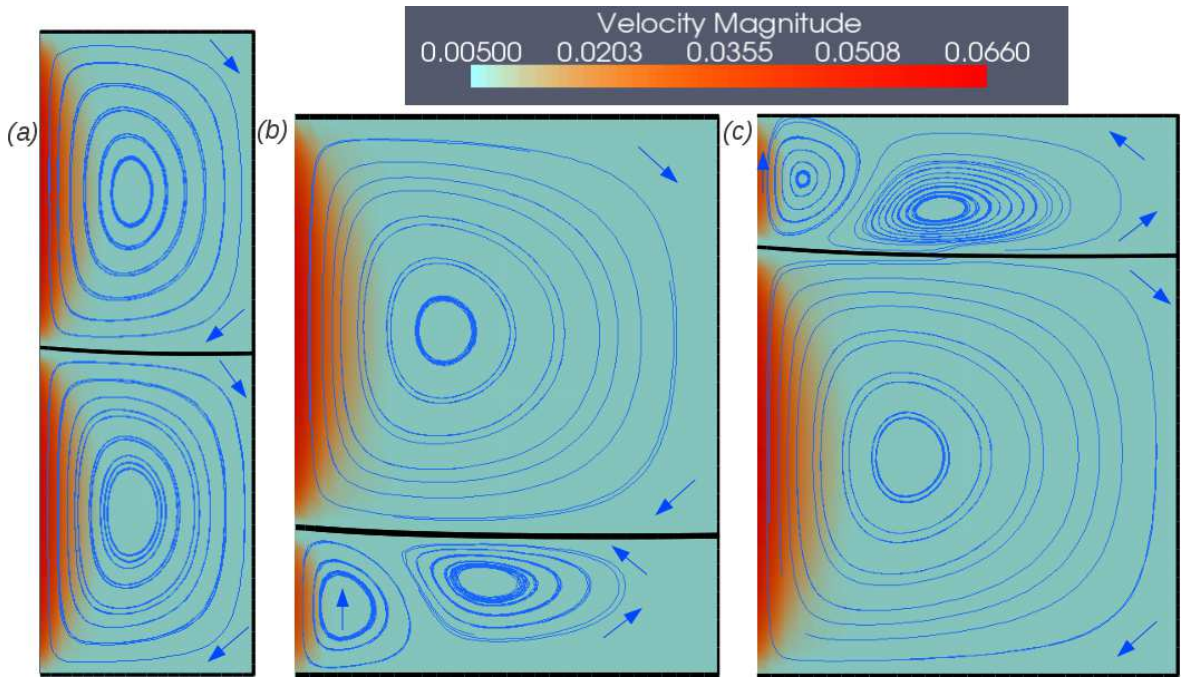}
\caption{Steady flow patterns. (\textit{a}) $R/L_1=2/3$  and $L_1/\omega_0=L_2/\omega_0=30$  (\textit{b}) $R/L_1=3$, $L_1/\omega_0=10$ and  $L_2/\omega_0=30$ (\textit{c}) $R/L_1=1$, $L_1/\omega_0=30$ and  $L_2/\omega_0=10$. In all cases $f_1/\sigma=0.3$, $\Sigma=f_1/f_2=1$, $\zeta=\eta_2/\eta_1=1$ and $Bo=10$.  Velocity magnitude is reduced by $U_\sigma$.}\label{box}
\end{center}
\end{figure}

The flow patterns in wide containers ($R>L_1,L_1=L_2$) and long containers ($R<L_1,L_1=L_2$) are displayed in figure \ref{2fluids} and figure \ref{box}(\textit{a}).\\
We first observe in figure \ref{2fluids}, nearly circular streamlines, which are the signature of toroidal flows also called eddies. These eddies exist at steady state because (i) the flow has to follow tangentially the interface 
due to the continuous axial forcing of the scattering force and (ii) mass conservation induces a circulation of the fluid.
 When $R>L_1$, figure \ref{2fluids} clearly shows that the spatial extension of these eddies scales like $L_1$ as discussed in the previous sections (figure \ref{vit-R-L}).
 When $L_1>R$ (figure \ref{box}(\textit{a})), the eddies invest the whole thickness of the layers 
while their width equals the container radius. To conclude, these results demonstrate that the axial dimension of the 
eddies is always $L_1$ while the radial one scales like the smallest dimension of the fluid layer.\\ 

Figures \ref{box}(\textit{b-c}) display the flow pattern when $L_1$ and $L_2$ are different.
 We notice first that the maximum velocities are different in the layers despite an equal scattering force density. This is due to the dependence of the velocity with respect 
to the thickness of the fluid layer, which has been evidenced in the left inset of figure \ref{vit-L-w0}.
 It should be also noticed that in the fluid layer of aspect ratio $L_1/L_2$ equal to $1$,
 there is an eddy extending to the whole layer, while in the other layer of smaller height, a secondary
 eddy develops. 
In fact, because the eddy is larger in the fluid layer of largest thickness, it forces a motion in the other layer, as a result of momentum transfer by viscous shear through the interface, yielding to a secondary eddy.
The scattering force exerted in both layers inducing two contrarotative eddies, any loss of symmetry of the flow domain induces the appearance of a secondary corotative eddy within one of the layers.
 Similar patterns can be obtained by imposing non equal scattering force densities in the two layers as illustrated in the next section.\\
\subsubsection{Influence of the scattering force ratio}
\begin{figure}
\begin{center}
 \includegraphics[scale=0.6]{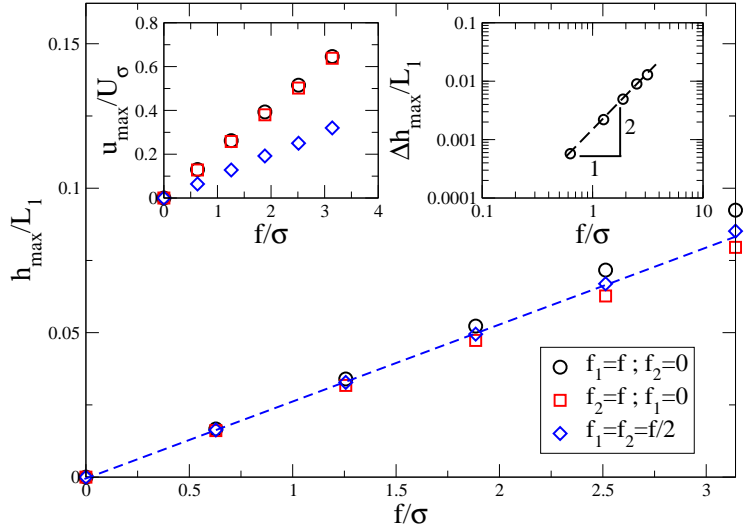}
\caption{Hump height variation against the total scattering force $f$ when $[f_1=f,f_2=0]$, $[f_1=0,f_2=f]$ and
 $f_1=f_2=f/2$ ($R/L_1=3.33$, $L_1/\omega_0=L_2/\omega_0=30$, $Bo=45$ and $\eta_1=\eta_2$). The dashed line is a linear regression for $f_1=f_2=f/2$. 
The left inset presents the maximum fluid velocity in each configuration while the right inset shows the variation of 
the reduced difference of heights $\Delta h_{max}/L_1$ between the cases $[f_1=f,f_2=0]$ and $[f_1=0,f_2=f]$ against $f/\sigma$.}\label{intensity}
\end{center}
\end{figure}

\begin{figure}
\begin{center}
 \includegraphics[scale=0.38]{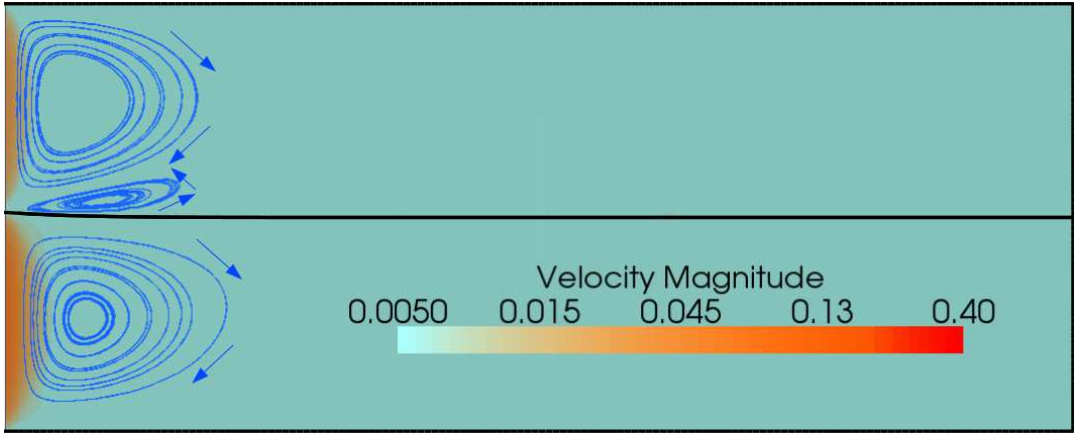}
\caption{Steady flow pattern for $R/L_1=5$ and $L_1/\omega_0=L_2/\omega_0=30$, $\Sigma=f_1/f_2=2$, $\zeta=\eta_2/\eta_1=1$, $Bo=10$ and $f_1/\sigma=0.3$. Velocity magnitude is reduced by $U_\sigma $.}\label{2fluidsdemi}
\end{center}
\end{figure}

\begin{figure}
\begin{center}
 \includegraphics[scale=0.385]{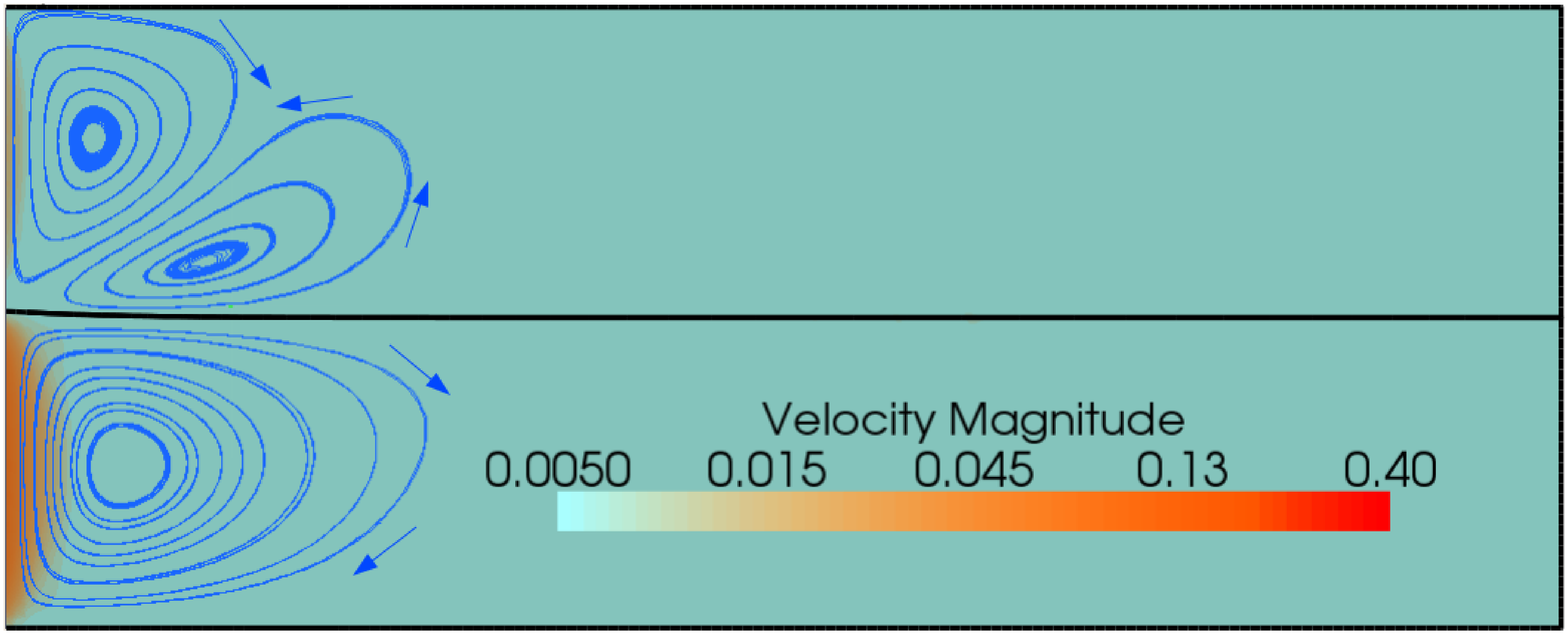}
\caption{Steady flow pattern for $R/L_1=5$ and $L_1/\omega_0=L_2/\omega_0=30$, $\Sigma=f_1/f_2=4$, $\zeta=\eta_2/\eta_1=1$, $Bo=10$ and $f_1/\sigma=0.3$. Velocity magnitude is reduced by $U_\sigma $.}\label{2fluidsquart}
\end{center}
\end{figure}

\begin{figure}
\begin{center}
 \includegraphics[scale=0.39]{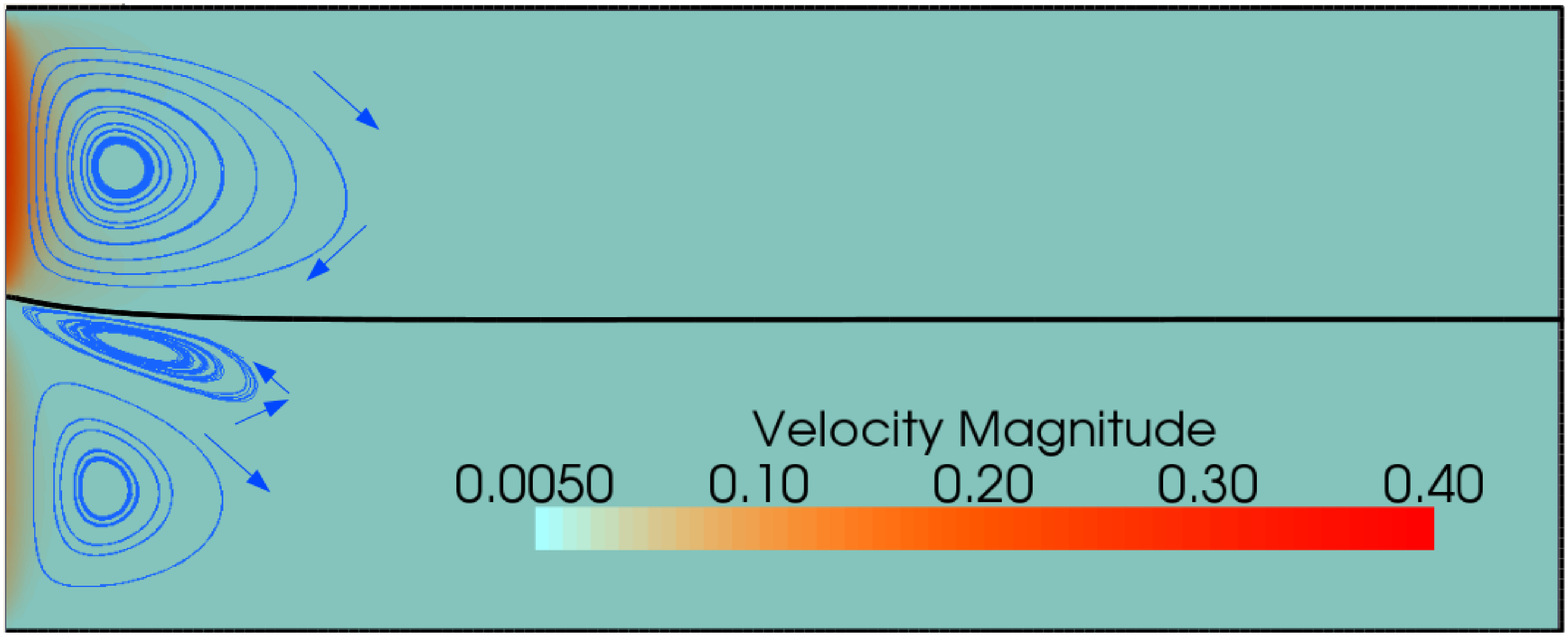}
\caption{Steady flow pattern for $R/L_1=5$ and $L_1/\omega_0=L_2/\omega_0=30$, $\Sigma=f_1/f_2=1/4$, $\zeta=\eta_2/\eta_1=1$, $Bo=10$ and $f_1/\sigma=0.3$. Velocity magnitude is reduced by $U_\sigma $.}\label{2fluidsquartinv}
\end{center}
\end{figure}

\begin{figure}
\begin{center}
 \includegraphics[scale=0.42]{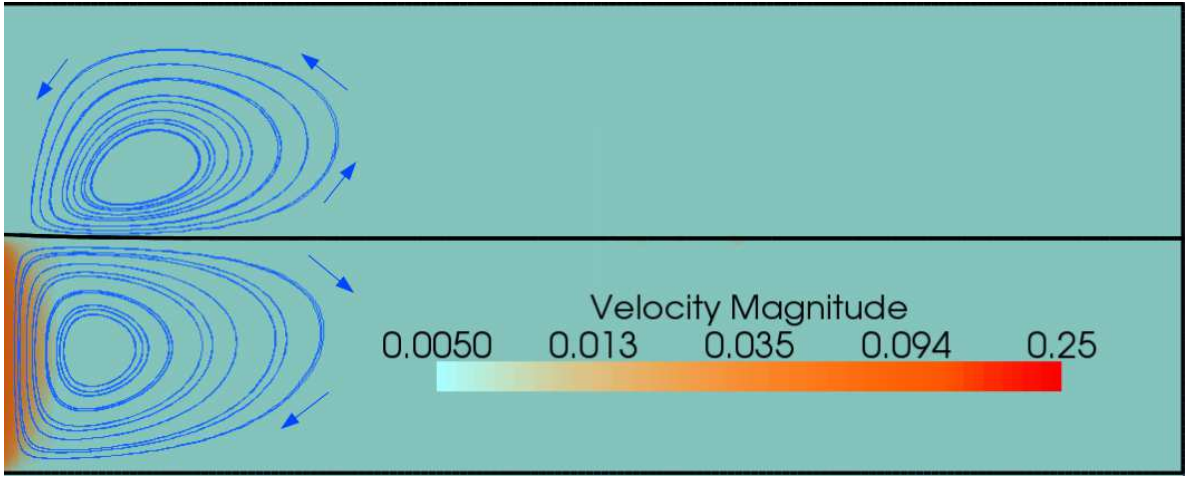}
\caption{Steady flow pattern for $R/L_1=5$ and $L_1/\omega_0=L_2/\omega_0=30$, $f_1/\sigma=0.3$, $f_2=0$, $\zeta=\eta_2/\eta_1=1$, $Bo=10$. Velocity magnitude is reduced by $U_\sigma $.}\label{onefluid}
\end{center}
\end{figure}

In the present section, we discuss the dependence of the flow and of the interface deformation on the magnitude of the scattering force.\\
Figure \ref{intensity} shows the variation of $h_{max}=h(r=0)$ with the total scattering force $f=f_1+f_2$. We first notice that when
 the forces are equal in both layers $f_1=f_2=f/2$, the variation of $h_{max}$ is linear with
 respect to $f$.
 For $[f_1=0,f_2=f]$ and $[f_1=f,f_2=0]$, we observe a small deviation from the linear behaviour, which is attributed to 
 the dependence of $h_{max}$ to the layer thickness as shown in figure \ref{int-L-w0}. Considering the case $[f_1=0,f_2=f]$, the layer thickness above the hump
 ($L_0-h_2$) is a little smaller than $L_0$ leading to a smaller deformation, while in the case $[f_1=f,f_2=0]$,
 the layer thickness below the hump ($L_0+h_1$) is a little larger 
 than $L_0$ leading to a larger deformation (see left inset of figure \ref{int-L-w0}). The right inset of figure \ref{intensity} shows that the variation of the difference of heights $\Delta h=h_1-h_2$ between
 the case $[f_1=f,f_2=0]$ and the case $[f_1=0,f_2=f]$ is quadratic with respect to $f$.\\
In the left inset of figure \ref{intensity}, we can see that the maximum velocities vary linearly with $f$,
 and that the same velocity is obtained in both configurations ($[f_1=f,f_2=0]$ and $[f_1=0,f_2=f]$).
 In the configuration $f_1=f_2=f/2$, the maximum velocity in each layer is half that obtained in the other configurations.
Consequently, the effects of the flows on interface deformations are additive as each layer induces its own viscous stress on the interface. \\
Considering the flow pattern, we observed in figure \ref{2fluids} that when $\Sigma=f_1/f_2=1$, two co-rotating eddies are induced,
 both of them rotating clockwise, as a result of the direction of the scattering force density along the $z$ axis. 
In this perfectly symmetric case, no secondary eddy is generated.\\
In figures \ref{2fluidsdemi}, \ref{2fluidsquart} and \ref{2fluidsquartinv}, corresponding to $\Sigma \neq 1$, a secondary eddy rotating counter-clockwise develops in the
 fluid layer where the amplitude of the scattering force is the smallest. This secondary eddy,
 whose extension increases with $\Sigma$, is contra-rotative compared to the main eddies.\\
Finally, in figure \ref{onefluid}, corresponding to $[f_1=f,f_2=0]$, we observe two contra-rotating eddies, the one in the bottom layer being determined
 by the direction of propagation of the laser beam while the one in the top layer,
 which is of weaker intensity, is exclusively induced by viscous shear through the interface as a result of zero scattering force density, $f_2=0$.\\
To summarize, when $f \lesssim \sigma$, there is no difference whether the scattering force is applied in the top or in the bottom layer.
 When $f > \sigma$, a small difference between the deformations can be noticed and is attributed to the resulting difference of thickness of the scattering layer.
 The flow patterns showed that a difference of scattering force between the layers leads to the emergence of a secondary eddy in the layer where the force is the smallest.\\
Up to now, we have investigated flow patterns for $\zeta=\eta_2/\eta_1=1$ in order to analyze the influence of the scattering force amplitude and of the different aspect ratios of the container.
 However, since viscosity affects flow velocities, the question remains on whether it modifies the two-layer flow pattern driven by light scattering.
This is the purpose of the following section.\\

\subsubsection{Influence of the viscosity ratio}
\begin{figure}
\begin{center}
 \includegraphics[scale=0.5]{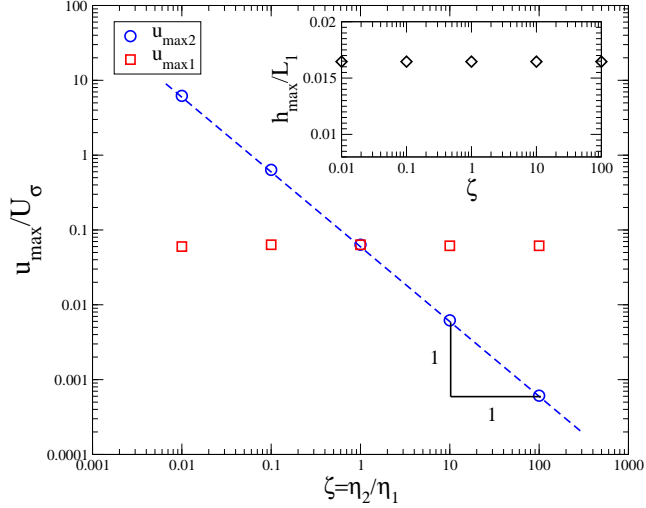}
\caption{Variation of the maximum velocities in each fluid layer as a function of the viscosity ratio $\zeta=\eta_2/\eta_1$.
 $\eta_2$ is varied whereas $\eta_1$ is kept constant ($\Sigma=f_1/f_2=1$, $f_1/\sigma=0.3$, $R/L_1=3.33$, $L_1/\omega_0=L_2/\omega_0=30$, $Bo=45$). The inset shows the variation of $h_{max}/L_1$ with
 $\zeta$.}\label{viscosity}
\end{center}
\end{figure}

In figure \ref{viscosity}, the variations of the maximum velocity $u_{max~i}$ in fluid layer $i$ is represented versus $\zeta=\eta_2/\eta_1$ while keeping $\Sigma=f_1/f_2=1$ and $L_1=L_2$.\\ As we set $\eta_1$ to a constant value,
 there is no variation of the velocity magnitude in 
layer $1$, while the velocity in layer $2$ scales like $1/\zeta=\eta_1/\eta_2$. This demonstrates that the dimensional scaling law established for a single layer flow $u_i\propto F_i  \omega_0^2/\eta_i $ ($f_i=F_i\pi\omega_0^2$) is valid in each fluid layer as
 $\omega_0$ is a characteristic length for the axial velocity near the $z$-axis (see comments of figure \ref{vit-L-w0}).
 Moreover, the inset of figure \ref{viscosity}
 shows that $h_{max}$ does not depend on the viscosity ratio.
 This can be explained by recalling that $T_{zzi}$ is proportional to $\eta_i u_i$ while $u_i$ is proportional to $1/ \eta_i$ making $T_{zzi}$ and hence $h_{max}$ independent of the fluid viscosities.\\
\begin{table}
\begin{center}
\begin{tabular}{|c|c|c|c|}
   \hline
$\zeta$  $\diagdown\diagdown$  $\Sigma$  & 0.1  & 1 & 10 \\  
   \hline
0.1  & 1  & 0 & 2 \\
 1   & 1  & 0 & 2\\
 10  & 1  & 0 & 2\\
   \hline
 \end{tabular}
 \end{center} \caption{Secondary eddy location for $(\Sigma,\zeta)\in\lbrace0.1;1;10\rbrace^2$. 
Bottom layer, top layer or inexistent secondary eddy are respectively marked as $1$, $2$ and $0$.}\label{table}
\end{table}
In order to get a general understanding of the location of the secondary eddy, a parametric study of the flow pattern was performed by varying $\Sigma$ and $\zeta$ in the set  
$(\Sigma,\zeta)\in\lbrace0.01;1;100\rbrace^2$ while keeping the total scattering force $f=f_1+f_2$ constant. Table \ref{table} summarises these results, indicating whether the secondary eddy is in the bottom layer (1), 
top layer (2) or nonexistent (0).\\
Table \ref{table} shows that the location of the secondary eddy is independent of $\zeta$ and
 is always located in the layer were the scattering force density is the smallest. 
This can be explained by noting that, since $F_1 \propto \eta_1 u_1/\omega_0^ 2$ and $F_2 \propto \eta_2 u_2/\omega_0^ 2$, one has 
 $u_1/u_2 \propto \Sigma\zeta$. This scaling law is confirmed by our computational results, as illustrated by the two following cases.\\
In the first case, $\Sigma=10$ and $\zeta=0.1$, we found that the maximum velocities in layers $1$ and $2$ are $u_{max~1}/U_\sigma \simeq u_{max~2}/U_\sigma \simeq0.012$. In the second case, $\Sigma=0.1$ and $\zeta=0.1$, we found $u_{max~1}/U_\sigma \simeq 0.00114$ and $u_{max~2}/U_\sigma \simeq 0.114$,
 in agreement with the prediction $u_1/u_2=0.01$.
The viscosity ratio only determines the velocity magnitude in each fluid layer.
This observation is confirmed by a simple model of parallel flow which is presented in the following section.\\
\subsubsection{Lubrication model}
The location of the secondary eddy is predicted by making use of a lubrication model assuming $R>>L_i>>\omega_0$ and derived from a two-dimensional 
cartesian formulation of the Stokes equation.
For the sake of simplicity, we consider a two-dimensional flow of velocity ${\boldsymbol u}=(u_x,0,u_z)$ in cartesian 
coordinates $(x,y,z)$. 
Since we are only interested in the distribution of eddies i.e. in the $z$ dependence of $u_x$,
 we consider layers of infinite horizontal extent and a parallel flow ${\boldsymbol u}=u_x(z){\boldsymbol e_x}$.
 Therefore far enough from the $z$-axis along which the forcing is exerted ($x>>0$), we can write the Stokes equation in both layers,  as follows
\begin{equation}
(a)~{0} = - \frac{\partial p_i}{\partial x} + \eta_i \frac{\partial^2 u_{ix}}{\partial z^2} ~~~;~~~\\
(b)~{0} = - \frac{\partial p_i}{\partial z} ~;~i=1,2
\label{stokes1d}
\end{equation}
We can deduce from (\ref{stokes1d} (b)) that the flow is driven by a horizontal pressure gradient.\\
Differentiating equation (\ref{stokes1d} (a)) with respect to $z$, we find that
\begin{equation}
{0} = \frac{\partial^3 u_{ix}}{\partial z^3} ~;~~i=1,2~~~\\
\label{D3u}
\end{equation}
This leads to the following expression of the velocity in each layer
\begin{equation}
u_{ix}(z)=a_iz^2+b_iz+c_i  ~;~~i=1,2~~~\\
\label{ux}
\end{equation}
where $a_i,b_i,c_i$, $i=1,2$ are six numerical constants to be determined by using the six following conditions
\begin{equation}
u_{1x}(z=0)=u_{2x}(z=0)~\\
\label{c1}
\end{equation}
\begin{equation}
\eta_1 \frac{d u_{1x}}{dz}(z=0)=\eta_2 \frac{d u_{2x}}{dz}(z=0)~\\
\label{c2}
\end{equation}
\begin{equation}
\int_{-L_1}^0 u_{1x}(z)dz=0~\\
\label{c3}
\end{equation}
\begin{equation}
\int_{0}^{L2} u_{2x}(z)dz=0~\\
\label{c4}
\end{equation}
\begin{equation}
\frac{d p_1}{dx}=-F_1~\\
\label{c5}
\end{equation}
\begin{equation}
\frac{d p_2}{dx}=F_2~\\
\label{c6}
\end{equation}
The first two conditions (\ref{c1}) and (\ref{c2}) simply express the continuity of velocities and shear stress
 along the interface. Conditions (\ref{c3}) and (\ref{c4}) express mass conservation in each layer. 
Conditions (\ref{c5}) and (\ref{c6}) express the fact that the pressure gradient 
driving the flow is related to the scattering force driving the motion near the axis as explained hereafter.
Because the eddies always scale like $L_i$ when $R>>L_i$, we consider that $\frac{\partial p_i}{\partial x} \propto \frac{\partial p_i}{\partial z} \propto p_i /L_i$.
In addition, the Stokes equation near the $z$-axis leads to $\frac{\partial p_i}{\partial z} \propto \eta_i \frac{\partial^2 u_{iz}}{\partial x^2} \propto F_i$ and finally to $\frac{\partial p_i}{\partial x} \propto p_i /L_i \propto F_i$.
One can also notice a (-) sign on equation (\ref{c5}). This is justified by the fact that when 
the system is symmetric (i.e. $F_1$=$F_2$, $\eta_1=\eta_2$, $L_1=L_2$), we assume that there is no motion of
 the interface along the $x$ axis, which leads to opposite pressure gradients and velocities near
 the interface ($u_{1x}(z=0^-)>0$ and $u_{2x}(z=0^+)< 0$). The solution approximated the flow without satisfying the no-slip boundary conditions at the top and bottom walls. The conditions imposed to the two-fluid, parallel flow lead to
\begin{equation}
c_1=c_2=\frac{F_1}{6\eta_1}L_1^2-\frac{L_1(\frac{F_1L_1^2}{3\eta_1}+\frac{F_2L_2^2}{3\eta_2})}{2(L_1+L_2/\zeta)}~\\
\label{c1b}
\end{equation}
\begin{equation}
 b_1= \zeta b_2=-\frac{\frac{F_1L_1^2}{3\eta_1}+\frac{F_2L_2^2}{3\eta_2}}{L_1+L_2/\zeta}~\\
\label{c2b}
\end{equation}
\begin{equation}
a_1=-\frac{F_1}{2 \eta_1}~\\
\label{c5b}
\end{equation}
\begin{equation}
a_2=\frac{F_2}{2 \eta_2}~\\
\label{c6b}
\end{equation}
\begin{figure}
\begin{center}
 \includegraphics[scale=0.7]{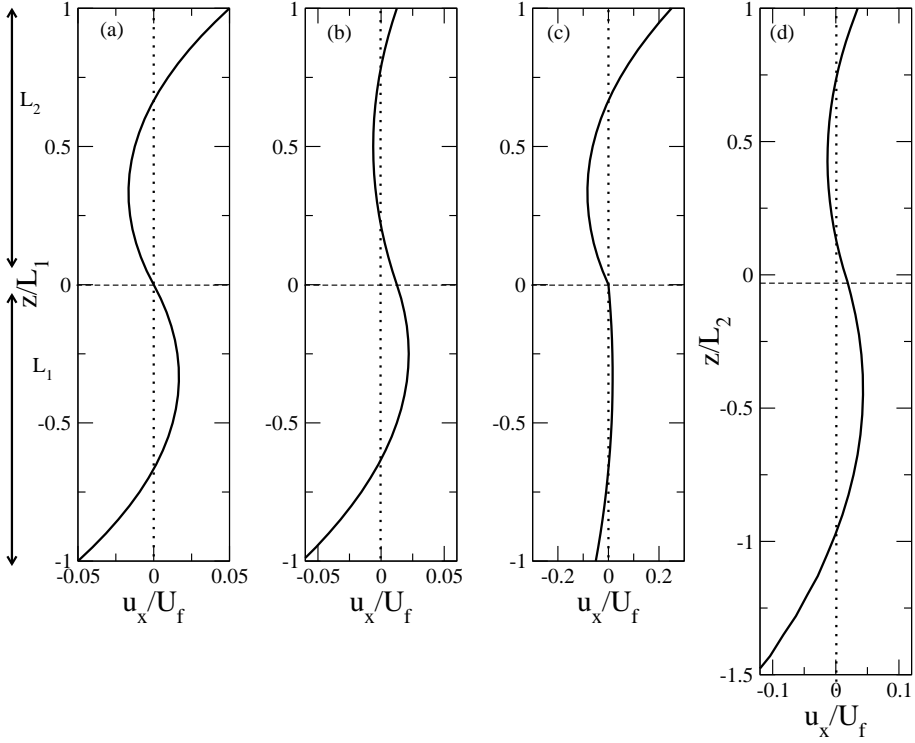}
\caption{Velocity $u_x(z)$ predicted by the cartesian model. (\textit{a}) Symmetric case: $\Sigma=1$, $\zeta=1$ and $L_1/L_2=1$. 
(\textit{b}) Contrast of magnitude of scattering force densities: $\Sigma=2$, $\zeta=1$ and $L_1/L_2=1$. (\textit{c}) Contrast of viscosities: $\Sigma=1$, $\zeta=0.2$ and $L_1/L_2=1$. 
(\textit{d}) Contrast of layer thicknesses: $\Sigma=1$, $\zeta=1$, $L_1/L_2=1.5$.}\label{1d}
\end{center}
\end{figure}
Figure \ref{1d} illustrates four interesting cases highlighting the emergence or not of a secondary eddy in one of the layers.
In figure \ref{1d}(\textit{a}), the system is symmetric and $u_x(z)$ changes sign only once in each layer. 
This means that only one eddy exists in each layer. In case (\textit{b}), the magnitude of the scattering force density in the top layer is half 
that in the bottom one. In the top layer, $u_x(z)$ changes sign twice (at $z=0.79$ and $z=0.21$). This means
 that two contra-rotative eddies exist in the top layer. In case (\textit{c}), the fluid in the bottom layer is 5
 times more viscous than the one in the top layer. We can notice that although the velocities in the bottom layer are much smaller,
 there is no secondary eddy. 
Finally, in case (\textit{d}) the bottom layer is $1.5$ times thicker than the top layer. A secondary flow appears in the layer of smallest thickness, since $u_x(z)$ changes sign twice at $z=0.73$ and $z=0.14$.
As a conclusion, even though very simple, this parallel flow model gives a qualitative prediction of the location of the
 secondary eddy confirming the numerical simulations of the axisymmetric flow.\\
\section{General discussion}
Optical streaming represents a new way to drive fluid flows by light. It can be useful in situations where the direction and magnitude of fluid flows need to be controlled.
The only condition for inducing these optical streaming flows is to use non absorbing turbid fluids. Any viscous fluid can be considered, while the strength of the optical streaming will strongly depend on 
the size and nature of the scatterers which could be solid or liquid nanosuspensions (see \cite{wunenburger10,schroll07} for an example of density fluctuations
 in near-critical binary mixtures acting as liquid scatterers) and generally speaking any fluid presenting refractive index inhomogeneities at a length scale much smaller than the optical wavelength.\\
Microfluidics is of particular interest for application of optical streaming since the small dimensions involved are compatible with the laser focusing.
 If a fluid-fluid interface is present in these experiments (either microfluidics coflow experiments or digital microfluidics), we expect the emergence of secondary eddies depending
 on the thicknesses of the fluid layers and on the scatterers present in these fluids. In addition, if the interfacial tension
 is sufficiently low, this could also lead to interface deformation. While a viscosity contrast will not influence the location of secondary flows,
 a significant impact of this contrast on flow magnitude is expected since velocity is inversely proportional to viscosity. Many applications of optical streaming can thus be devised.
 For instance, flow control at very small flow rates is always difficult
 in microchannels as syringe pumps operate close to their mechanical limits; flow rate changes are also always accompanied by complex transients. In these regimes, the fine and smooth control
 of viscous flows by optically tunable scattering forces is probably an interesting alternative for turbid fluids. Micromixing is also very challenging in classical microchannels \citep{nguyen05}
  because flows are mostly laminar and thus constitute a major drawback for efficient chemical reactions \citep{baroud03}. One route for active mixing consists in superimposing pressure
 perturbations to the main stream \citep{lee01b}. Scattering forces could replace pressure actuation if fluids are turbid or seeded with appropriate scatterers.\\ 
\section{Conclusion}
In this work, we investigated flow production by light scattering in a two-fluid system involving turbid fluids.
 We compared the flow pattern predicted numerically to the analytical one-fluid model prediction. 
This simplified analytical description predicts most of the characteristic features of both the induced flow and the resulting interface deformation.
Due to the large scale of induced flows, we investigated the geometrical effects  
of the container. We showed that the velocity field and the interface deformation scale like the smallest dimension of the container.
 Moreover a variety of flow patterns were illustrated  when breaking the system symmetries.
 For instance, in absence of scattering force intensity contrast, secondary eddies always emerge in the thinnest layer
while in absence of layer thickness contrast, secondary eddies emerge in the fluid layer where the scattering force density is the weakest.
On the contrary the variation of the viscosity ratio only affects velocities within
 the fluids while neither the interface deformation nor the flow pattern at steady state are influenced. Finally, a lubrication model was used to explain the emergence and to predict the location of these secondary flows in the layers.\\
Generally speaking, light-induced flows exist whenever isotropic scattering fluids are at work; 
fluids near a second-order phase transition can be an interesting example to perform experiments. 
Consequently, we have proposed and well-identified a general non dissipative way to produce bulk flow with light
 that goes one step forward in the exploration of the coupling between light and hydrodynamics, so called ''optohydrodynamics''. 
\section{Appendix : Comparison with experiments : Interface deformation in the presence of radiation pressure}
\begin{figure}
\begin{center}
 \includegraphics[scale=0.7]{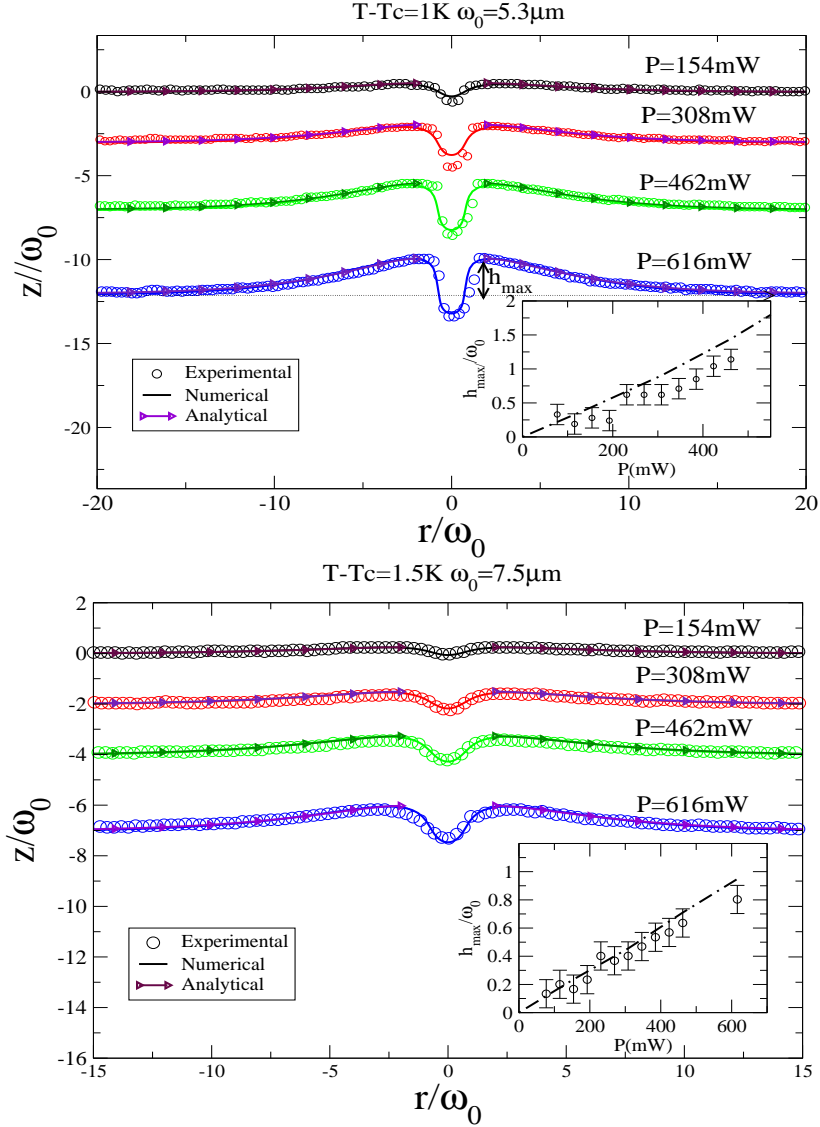}
\caption{Comparison of steady state interface profiles obtained from experiments, analytical solutions and numerical simulations. Top: T-T$_c$=1 K, $\omega_0$=5.3 $\mu$m,
 Bottom: T-T$_c$=1.5 K, $\omega_0$=7.5 $\mu$m. Insets show the variation of scattering deformation $h_{max}/\omega_0=h(r/\omega_0=2)$ as a function of beam power $P$ 
(dot-dashed line: numerical, symbols: experiments). 
In all cases the optical wave propagates upward. Interface profiles were shifted vertically for clarity.}\label{profils}
\end{center}
\end{figure}

In this appendix, we show comparisons between numerical, analytical and experimental steady profiles of interface deformations.
In experiments involving turbid two-phase systems like the one used in \cite{schroll07}, the interface is made visible thanks to a refractive index contrast.
In addition to the hydrodynamics stress due to the optical streaming, radiation pressure is exerted on the interface (see \cite{chraibi10}). In this appendix, both stresses are taken into account in the numerical simulation.
Note that in \cite{schroll07}, only optical streaming effects were taken into account while in \cite{chraibi10}, only radiation pressure effects were modeled.\\
In order to model radiation pressure effects, we have to add the following term (which is presented in details in \cite{chraibi10}) in the right side of equation (\ref{stressjump}) 
\begin{equation}
\Pi (r)=\frac{I(r)}{c}n_2\cos \theta _{i}(2(1-\delta)\cos \theta
_{i}- \Upsilon(\theta _{i},\theta _{t}) ((1-\delta)\cos \theta _{i}+\cos \theta _{t}))
\label{radup}
\end{equation}
The radiation pressure is due to the jump in photon linear momentum (proportional to the refractive index $n_i$) 
when propagating from fluid 1 to fluid 2.
$\Upsilon(\theta _{i},\theta _{t})$ is the Fresnel
transmission coefficients of energy fluxes for circularly polarized beams \citep{chraibi08a}
\begin{equation}
\Upsilon(\theta _{i},\theta _{t})=\frac{2(1-\delta)\cos \theta _{i}\cos \theta
_{t}}{((1-\delta)\cos \theta _{i}+\cos \theta _{t})^{2}}+\frac{%
2(1-\delta)\cos \theta _{i}\cos \theta _{t}}{(\cos \theta
_{i}+(1-\delta)\cos \theta _{t})^{2}}
\end{equation}
where $\theta_i$ and $\theta_t$ are the incidence and transmission angles at the fluid interface respectively. They are such that $\theta_i=\arctan(\frac{d h}{dr})$ and $\theta_t=\arcsin((1-\delta) \sin\theta_i)$
 where $\delta=(n_2-n_1)/n_2$ is the relative refractive index contrast.\\
The fluid used in this experiment is a near-critical binary mixture in the two-phase regime that undergoes refractive index fluctuations
of characteristic size $\xi^-$, which increases as the critical point is neared. When the
 difference between the sample temperature T and the critical temperature T$_c$ is large,
 $\xi^-$ vanishes and each phase at coexistence can be considered as almost homogeneous \citep{casner01a,casner03a} for the light-matter interaction.
 However, when T is close to T$_c$, $\xi^-$ increases up to tens of nanometers at T-T$c$=0.1K and the sample significantly scatters light, producing strong linear momentum transfer to the fluid.
Details on the experiments will not be presented here, however one can refer to \cite{schroll07}, \cite{wunenburger10} and \cite{chraibi10} for more details.\\
The profiles are analytically predicted using the stream function formulation developed in section 3 all along with the numerical solution of equation (\ref{equi}) (considering $h'(r=2\omega_0)=0$ as observed in the experiments).\\
Figure \ref{profils} presents comparisons between two-fluids numerical results, one-fluid analytical predictions and experimental steady interface
 profiles for two different experimental conditions (T-T$_c$=1 K, $\omega_0$=5.3 $\mu$m,
 and T-T$_c$=1.5 K, $\omega_0$=7.5 $\mu$m) and at four different values of the laser beam power.\\ 
A very good agreement is obtained between the experimental data and the predictions at each power value $P=154,$ $308,$ $462$ and $616$ mW. Insets of figure \ref{profils} illustrate the linear power dependence
 of the deformation height due to the steady scattering flows disturbed by opposite radiation pressure effects.\\ All these results validate both the numerical resolution and the theoretical model,
 which both represent important tools to describe these original physical phenomena.\\ 

\newpage
\bibliographystyle{jfm2}

\end{document}